\documentclass[12pt]{article}

\usepackage{graphicx,amssymb,url,mathrsfs, amsmath}
\usepackage{eucal}
\usepackage{wrapfig}
\usepackage{boxedminipage}
\usepackage{setspace}
\usepackage{subfigure}

\def\IR{{\hbox{{\rm I}\kern-.2em\hbox{\rm R}}}}
\def\IB{{\hbox{{\rm I}\kern-.2em\hbox{\rm B}}}}
\def\IN{{\hbox{{\rm I}\kern-.2em\hbox{\rm N}}}}
\def\IC{\,\,{\hbox{{\rm I}\kern-.59em\hbox{\bf C}}}}
\def\IZ{{\hbox{{\rm Z}\kern-.4em\hbox{\rm Z}}}}
\def\IP{{\hbox{{\rm I}\kern-.2em\hbox{\rm P}}}}
\def\IH{{\hbox{{\rm I}\kern-.4em\hbox{\rm H}}}}
\def\ID{{\hbox{{\rm I}\kern-.2em\hbox{\rm D}}}}

\def\be{\begin{equation}}
\def\ee{\end{equation}}
\def\ba{\begin{eqnarray}}
\def\ea{\end{eqnarray}}

\def\ea{{\it et al}. }

\begin{document}

\begin{titlepage}

\vspace{0.5in}

\begin{center}
{\large \bf Holographic dual of Cold Trapped Fermions}\\
\vspace{10mm}
Sang-Jin Sin$^{a}$ and Ismail Zahed$^{b}$\\
\vspace{5mm}
{\it $^b$ Department of Physics, Hanyang University,  Seoul Korea\\
$^a$ Department of Physics and Astronomy, Stony Brook University, Stony Brook NY 11794}\\
          \vspace{10mm}
  {\tt \today}
\end{center}
\begin{abstract}
We study cold fermionic atoms using the holographic principle.
We  note that current atomic experiments with massive fermions trapped
in a harmonic potential in the unitarity limit  behave  as massless fermions  thanks to the Thomas-Fermi approximation.  We  map the thermodynamics of strongly correlated massless fermion to
that of the  charged black hole and study the thermodynamics and 
transport properties of cold fermions at strong coupling at finite temperature and 
density. In cold limit, the specific heat of charged black hole  is linear in $T$   independent of the dimensionality, which is reminiscent of Fermi  liquids. The shear viscosity per particle is shown to be finite
as a consequence of the non-vanishing of the entropy. 
We show that our holographic results compare favorably with most of the
current atomic data.  
\end{abstract}
\end{titlepage}


\renewcommand{\thefootnote}{\arabic{footnote}}
\setcounter{footnote}{0}



\section{Introduction}

Recent atomic measurements involving trapped cold fermions have unravelled a wealth
of results regarding the behavior of dilute but strongly interacting fermionic systems
\cite{ATOMS}.  By varying the strength of the magnetic trap, atoms were driven through the
notorious Feshbach resonance making their scattering length for all purposes infinite.

Strongly interacting cold fermions with an infinite scattering length exhibit  universal bulk
and transport properties.  In particular, the ground state energy per particle was found to
asymptote $E/NE_F\approx 0.5$ independently of the details of the interaction \cite{UNIVERSAL}.
In the long wavelength limit, strongly interacting fermions behave hydrodynamically for
a broad range of temperatures above the superfluid temperature~\cite{HYDRO}.

It was suggested in~\cite{GELMAN}  that cold fermionic systems may exhibit  also universal
transport properties with in particular a quantum bound on the viscosity to density ratio with
$\eta/n\geq \hbar/6\pi$.  This bound is analogous but distinct from the quantum bound for hot
gauge theories established using the holographic principle \cite{HOLO}.  Indeed, transport in
cold systems is affected by the interactions through possibly a Fermi surface or a fermionic crystal
as opposed to transport in hot systems which is the result of rescattering in the heat bath. The cold
quantum bound on the viscosity was recently explored experimentally in~\cite{THOMAS}.

Finite fermion number as well as temperature is crucial to analyze strongly coupled
fermionic systems in the context of the gravity dual theories.
Finite fermion density in the holographic approach  was introduced in~\cite{FIRST} and has since been
discussed by many.  The gravity back reaction of the dense matter is also encoded by considering
bulk filling branes \cite{SIN}, which allows the charge of  the RN AdS black hole to be identified with the
fermionic charge. The hydrodynamics and transport coefficients  of this system have been analysed in \cite{Son1,SIN2},
where fermionic corrections at high temperature have been discussed.

In this paper,  we suggest that the RN AdS black hole can describe cold fermions as it interpolates
continuously between hot and cold massless fermions. We use dual gravity to study the bulk thermodynamics
and transport near the cold limit.  In the cold limit, the specific heat of charged black hole  is linear in $T$ independent of dimension like in Fermi liquids.
Notice that this is consistent with recent studies of fermionic propagators  in~\cite{lee,liu,zaanen} which report the  
the presence of the fermi surface at zero temperature. We also note that recently in~\cite{Son2,Adams} a  non-relativistic gravity background was suggested 
for cold atoms. Here we use the relativistic gravity background based on the observation that nonrelativistic atoms in harmonic traps behave similarly to massless fermions in the Thomas-Fermi approximation. The latter is well justified for current atomic traps with
about a million atoms at the coldest temperature.

In section 2, we
briefly review the holographic model. In section 3 we detail the   thermodynamical results
at finite temperature and density. In section 4 we analyze the cold limits of transport properties
and derive the longitudinal and transverse fermionic conductivities. In section 5, 
we interpret the results for zero temperature using the visco-elastic framework to derive the
constitutive equations for the longitudinal and transverse fermionic currents. In section 6, we show that
magnetically trapped massive fermions behave as free massless fermions in 3 spatial dimensions.
In section 7, we compare our results to a number of thermodynamic and transport results established recently
using cold and trapped fermions. We conclude with summary, a caution and prospects in section 8.

\section{Holographic Model}

To analyze cold fermions at low temperature we make use of a simple holographic model~\cite{SIN} in which fermions are included as a U(1) charge  source on the bulk filling branes and  the fermion back reaction on the  metric is included.   
 Although the original  model was motivated from the D3-D7 embedding, here we introduce it 
  as a  toy model describing the condensed matter system at the 
  boundary. The model  is the  Reissner-Nordstrom AdS (RN-AdS) black hole metric \cite{adsBH} coupled to a U(1) flavor bulk field  sourced by the fermion charge on the BH.    
 We take the RN AdS BH as a model in the bottom up approach. The global fermionic charge
 should be identified with a local U(1) gauge field in bulk. As we have only one U(1) vector gauge field  in bulk,  it is natural to identify it as the dual to the fermionic  charge on the boundary in our bottom up approach.   
  
  The thermodynamics \cite{adsBH} and transport \cite{Son1,SIN2} in the RN-AdS have been analyzed with a particular emphasis
  on density corrections to the high temperature limit~\cite{SIN}. Here instead, we will be mostly
  interested in the cold temperature limit of the RN-AdS results  as they may shed light on bulk thermodynamics and transport in cold and   non-confining fermionic systems at strong coupling as we detail below.   The interest in these systems is obvious from
  the wealth of atomic experiments that have sprung in the past decade using atomic trap experiments with cold fermions near the Feshbach resonance. 

  The effective action describing bulk RN-AdS gravity  sourcing a U(1) gauge field reads

  \be
  S=\frac 1{2\kappa^2}\int d^5x\sqrt{-g}(R-2\Lambda) -\frac 1{4e^2}\int d^5x\sqrt{-g}F^2
  \label{RNADS}
  \ee
  with $\kappa^2=8\pi G_5$ and $\Lambda=-6/l^2$ are the gravitational and cosmological
  constant~\footnote{ If we consider this model as stemming from a
  leading order approximation of the D3-D7 embedding, the U(1) gauge coupling is $l/e^2=N_cN_f/(2\pi)^2$  and 
  $l^3/\kappa^2=N_c^2/4\pi^2$.  In 10 dimensions  the bulk filling procedure is
  optimal for D9 rather than D7 branes.  For D9, $1/e^2=(N_cN_f/32\pi^3)l/l_s^2$
  and $l^3/\kappa^2=N_c^2/4\pi^2$.}.    
  At this stage, one may wonder how to map the gravity parameters to a system of 
 cold fermions. Our strategy is to express all the bulk quantities in terms of  $\kappa$ and the ratio $\gamma:= \kappa/le$. 
  As we  show below, $\kappa$  drops   in the thermodynamic {\it per particle} and $\gamma$ is mapped on a
  universal parameter of the cold fermionic systems.
  
  The metric for the RN-AdS black-hole is

  \be
  ds^2=\frac {r^2}{l^2}\left(-f\, dt^2+d\vec{x}^2\right)+\frac {l^2}{r^2 f}dr^2
  \label{MADS}
  \ee
  and the U(1) gauge field is

  \be
  A_4=\mu-\frac Q{r^2}
  \label{GA}
  \ee
  with $f=1-ml^2/r^4+q^2l^2/r^6$.  The gravitational equations of motion tie the electric charge
  $Q$ to the geometrical charge $q$ through

  \be
  \frac{q^2}{Q^2}=\frac 23 \frac{\kappa^2}{e^2}
  \label{CHARGE}
  \ee

 The position of the RN-AdS black-hole horizon is fixed  by the maximum zero of $f(r_+)=0$,
  which exists if and only if the geometrical charge $q$ satisfies the inequality $q^4\leq 4m^3l^2/27$.
  Again $m$ and $l$ are the mass and length of the RN-AdS black hole space,  with $r_+$ bounded by
  $l\sqrt{m/3}\leq r_+^2\leq l\sqrt{m}$.   The fermion chemical potential $\mu$ in (\ref{GA}) is fixed by
  demanding that $A_4(r_+)=0$ on the black-hole horizon. This latter condition enforces the Gibbs relations.
  The temperature in the RN-AdS space is defined by the singulaity free conical condition

  \be
  T=\frac{r_+^2f'(r_+)}{4\pi l^2}=1/\beta
  \label{TEMP}
  \ee

  The regulated Gibbs energy $\Omega=\Delta S/\beta$ follows from (\ref{RNADS}) by inserting
  the RN-AdS charged black hole (\ref{MADS}-\ref{GA}) and subtracting the {\it empty} thermal
  AdS contribution~\cite{SIN}. The result is

   \be
   \Omega=-\frac{V_3}{2\kappa^2l^3}\left(\frac{r_+^4}{l^2}+\frac{q^2}{r_+^2}\right)
   \label{OMEGA}
   \ee
   by trading $m=r_+4/l^2+q^2/r_+^2$. The densities of entropy  $s$, energy $\epsilon$ and pressure $p$
    follow from (\ref{OMEGA}) through the usual grand-canonical rule~\cite{adsBH,SIN}
   \begin{eqnarray}
   s=&&\frac{2\pi r_+^3}{\kappa^2l^3}=\frac{\pi l^3}{4\kappa^2b^3}\nonumber\\
   \epsilon=&&\frac{3m}{2\kappa^2l^3}=\frac{3l^3}{32\kappa^2b^4}(1+a)=3p
   \label{EOS}
   \end{eqnarray}
   with $a=q^2l^2/r_+^2$ and $b=l^2/2r_+$.
   The fermion chemical potential and number density are $\mu=Q/r_+^2$ and $n=2Q/e^2l^3$.

  \section{Thermodynamics at Low Temperature }

 The above thermodynamical relations hold for all values of temperature $T$ and chemical potential
 $\mu$ for fixed $\gamma=\kappa/le$.
 First, we examine the thermodynamics in the cold limit  to acquire more physical insights on the cold system.
  The fermions that source the U(1) vector fields  are expected to be in the Coulomb phase with strong vector
  interactions. At weak coupling 
 the particle-hole (ph) interaction
 is larger than the particle-particle (pp) interaction.
 However, small gaps and phase space rules in favor of the BCS pairing over the Overhauser  mechanism~\cite{OVERHAUSER}.   What happens at strong coupling is currently speculative,
 in the absence of first principle calculations. The current holographic model may provide some
 insights to this question.

Unwinding the above thermodynamical relations in terms of temperature and chemical potential yield
for the Gibbs potential

\be
\Omega/V_3=\mu n=\epsilon +p -Ts
\ee
Introducing the parameter $N_c$ through
\be
N_c := \frac{2\pi l^{3/2}}{\kappa} , 
\ee
the fermion number density is explicitly
\be
n=\frac{N_c^2 \gamma^2}{8}\mu T^2\,\left(1+\sqrt{1+\frac{4\gamma^2 \mu^2}{3\pi^2T^2}}\right)^2
\label{DENSITY1}
\ee
with an expansion around the cold point of the form

\be
n=\frac {N_c^2\gamma^4}{6\pi^2}\mu^3+ \frac{N_c^2\gamma^3}{2\pi\sqrt{3}}\mu^2T+
\frac{1}{4}  N_c^2\gamma^2 \mu T^2+
 \frac{\sqrt{3}\pi }{16}  N_c^2\gamma T^3+
{\cal O}(T^5)
\label{DENSITY2}
\ee
Note that the presence of linear term in (\ref{DENSITY2})  can be  readily seen from (\ref{DENSITY1}) when $T^2$ is
combined in the squared bracket, ie 
$$ \left(T+\sqrt{T^2+\frac{4\gamma^2 \mu^2}{3\pi^2}}\right)^2.$$
The RN-AdS black-hole at zero temperature is characterized by a finite fermion density which is proportional to $N_c^2\gamma^4$. 
This result is the analogue of the R-charge density in the D3 case with a U(1) vector potential,  ie $n=N_c^2\mu^3/96\pi^2$
with $\gamma =1/2$.
The small temperature correction is linear in $T$ and of order $N_c^2\gamma^3$. 

The energy density reads
\begin{eqnarray}
\epsilon=\frac{3\pi^2N_c^2T^4}{128}\left (1+\sqrt{1+\frac{4\gamma^2 \mu^2}{3\pi^2T^2}}\right)^4
\left(1+\frac{8N_f\mu^2}{3N_c\pi^2T^2} \left(1+\sqrt{1+\frac{4\gamma^2 \mu^2}{3\pi^2T^2}}\right)^{-2}\right)
\label{ENERGY1}
\end{eqnarray}
and its expansion around the cold fermionic point is

\be
\epsilon=\frac{N_c^2\gamma^4}{8\pi^2}\mu^4+\frac{N_c^2\gamma^3}{2\pi\sqrt{3}}\mu^3\,T+\frac{3 N_c^2\gamma^2}{8}\mu^2T^2
+{\cal O}((T/\mu)^3)
\label{ENERGY2}
\ee Note that the equation of state at zero 
temperature
\be
\epsilon=\frac 34 \left(\frac{6\pi^2}{\gamma^2N_c^2}\right)^{1/3}\,n^{4/3}
\ee
suggest that this is a system  of degenerate massless fermions.    
This dual interpretation is supported by transport properties
as we discuss below. These massless states are the analogue to the BPS marginal bound states in the supersymmetric theory.  
 
A similar take on the current results follows from the entropy through
\be
s=\frac{\pi^2 N_c^2T^3}{16}\left (1+\sqrt{1+\frac{4\gamma^2 \mu^2}{3\pi^2T^2}}\right)^3
\label{ENTROPY1}
\ee
and its expansion around the cold point
\be
s=\frac{N_c^2\gamma^3}{6\pi\sqrt{3}}\mu^3+\frac {N_c^2\gamma^2}{4}\mu^2T+\frac{3\sqrt{3}\pi N_c^2\gamma}{16}
\mu T^2 +{\cal O}(T/\mu)^3)
\ee
The entropy density is finite at zero temperature and proportional to $N_c^2\gamma^3\mu^3 $.
This cold entropy is the degree of degeneracy of the fermionic liquid.


The specific heat near the cold point is 
\be
c_V=T\frac{ds}{dT} =\frac {N_c^2\gamma^2}{4}\mu^2T+\frac{3\sqrt{3}\pi N_c^2\gamma}{8}
\mu T^2 +{\cal O}(T/\mu)^3)
\label{SPECIFIC}
\ee
 The low-temperature leading contribution in (\ref{SPECIFIC}) is linear in the temperature.  Notice that this is common to all RN AdS black hole independent of  dimension, which is very similar to the specific heat observed in ordinary Fermi liquids.     Note that
 \be
\frac{ N_c^2\gamma^2}{4}\mu^2T= \frac3 {\gamma^2} \cdot\left(\frac{\pi^2}2\frac{ nT}{ \mu }  \right) \ee
where the bracket is suggestive of the Fermi liquid density.  We recall that in the presence of a Fermi surface,
the specific heat feeds only from the fermions on the surface fluctuating in  its orthogonal  direction, thus $\mu^2 T$.

The ratios of the energy and entropy densities  to the fermion number density are respectively
\be
\frac{\epsilon}n=\frac{ \pi^2T^2}{16\gamma^2\mu}\left(\left(1+\sqrt{1+\frac{4\gamma^2 \mu^2}{3\pi^2T^2}}\right)^{2}
+\frac{8 \gamma^2\mu^2}{3 \pi^2T^2} \right)
\label{DENSITY3}
\ee
and
\be
\frac{s}{n}=\frac{ \pi^2T}{2\gamma^2\mu}\left (1+\sqrt{1+\frac{4\gamma^2 \mu^2}{3\pi^2T^2}}\right)
\label{ENTROPY3}
\ee
Their small temperature expansions are respectively

\be
\frac{\epsilon}{n\mu}=\frac 34 +\sqrt{\frac{3 }{4 }} \,\frac{\pi T}{2\gamma\mu}+{\cal O}((T/\mu)^2)
\label{RATIO34}
\ee
and

\be
\frac{s}{n}=\frac{\pi}{\sqrt{3}} \frac1{\gamma}+
\frac{\pi^2 }{2 \gamma^2}\frac T\mu +{\cal O}((T/\mu)^2)
\label{ENTROPYRATIO}
\ee

Here we comment on an interesting numerology which we do not 
understand fully. First, we note that at  zero temperature and finite density, 

\be
\frac{\epsilon}{n\mu}=\frac{E}{E+PV}=\frac{E}{E+E/3}=\frac 34
\label{3over4}
\ee
in general thanks to the first law of thermodynamics ( $E+PV=TS+\mu N$) and   conformal symmetry ($PV=E/3$).  Similarly, 
  at  zero density   
 
\be
\frac{\epsilon}{TS}=\frac{E}{E+PV}=\frac{E}{E+E/3}=\frac 34
\label{3over4T}
\ee
for exactly the same reasons.  
Now, we remind the readers that at high temperature the {\it interacting} $\epsilon$ and $p$ are also  3/4 the leading black-body contributions~\cite{KLEBANOV}. The same factor of 3/4 will also emerge from harmonically trapped cold atoms as we explain
below.  It is also interesting to notice  that  the leading density ($\mu$ dependent)  contributions  to $n, \epsilon, p$ at high temperature are 3 times the expected scalar and bifundamental fermion contributions.

Finally, the finite density construction followed in
\cite{SIN} and the present work treats the density and temperature on equal footings by means of the
RN-AdS black-hole. It is worthy to notice that the RN-AdS black hole horizon

\be
r_+=\frac{l^2}{2}\left (\pi T+\sqrt{\pi^2T^2+\frac{4\gamma^2\mu^2}{3}}\right)
\rightarrow  \frac{l^2}{\sqrt{3}}\gamma\mu
\label{RADIUS}
\ee
does not vanish at zero temperature. This radius is to be compared with
the one at finite temperature and zero density where  $r_+\rightarrow l^2 \pi T$. Interestingly enough,
the square root dependence on $T,\mu$ in (\ref{RADIUS}) is the one expected from QCD-like arguments
~\cite{QCD}. This point will be explored in a further investigation.
Here we see that the transition from dense to hot occurs for

\be
\pi T\approx 2\mu\gamma/\sqrt{3}
\ee
 
One intriguing part of the black-hole holography for cold fermions is the appearance of a finite {\it entropy}  per particle
in the zero temperature limit tied with the finite horizon radius. Namely, 
\be
\frac SN=\frac sn=\frac{\pi}{\sqrt{3}\gamma}=f.
\label{ENTROPYZERO}
\ee  This persistent cold entropy is at the origin of the viscosity of the superfluid mode in the cold atoms to be detailed below.

What about the third law? The Nernst theorem or the third law of the thermodynamics  asserts that the entropy at zero temperature 
approaches to the minimum value. The minimum value 
is almost always zero, however,   it is not  zero if the ground state is degenerate.  
The holographic approach
to strongly coupled fermion systems suggests that the  cold atoms have  small but
measurable entropy at zero temperature.  If true, it provides an example of non-vanishing residual entropy.  
Below we explore
the possible measurability of this entropy in cold fermionic atoms either through calorimetry or viscosity measurements in the nonsuperfluid phase near zero temperature.

\section{Transport Analysis}

Transport in cold holographic fermionic systems has been discussed recently in D4/D8
\cite{KIM} and D3/D7 \cite{SON}.  In this section we focus on the fermionic
response function for the RN-AdS black-hole construction.  Let ${\bf A}_\mu(x)$ be the source of the
{\it fermion} bilinear 4-vector current in the boundary of AdS$_5$ ($r=\infty$).  The  expectation value of the fermion current is

\be
{\bf J}_\mu(x)=-i\int d^4y\,\left< J_\mu(x)J_\nu(y)\right>_R \,{\bf A}^{\nu}(y)
\label{JJ}
\ee
in the linear response approximation. Here $R$ refers to the retarded correlation function
in the state of finite temperature and density.  In Fourier space, (\ref{JJ}) simplifies

\be
{\bf J}_\mu(k)=G^R_{\mu\nu} (k)\,{\bf A}^\nu(-k)
\label{RRF}
\ee
with the retarded Green's function

\be
G^R_{\mu\nu}(k)=-i\int\,d^4x\,e^{ik\cdot x}\left<J_\mu(x)J_\nu(0)\right>_R
\label{GREEN}
\ee
For a plane-wave baryonic 4-vector field ${\bf A}=(0,A^1,A^2,A^3)$ with wavenumber
$k^\mu=(\omega, 0,0,k)$ along the z-direction,  the induced current in (\ref{RRF})
can be decomposed into longitudinal (along $k$) and transverse (orthogonal to $k$)
components

\begin{eqnarray}
&&{\bf J}_L(k)=\left(\frac{G_{zz}(k)}{-i\omega}\right)\,{\bf E}_L(-k)\nonumber\\
&&{\bf J}_T(k)=\left(\frac{G_{xx}(k)}{-i\omega}\right)\,{\bf E}_T(-k)
\label{JJLT}
\end{eqnarray}
with the baryon electric field ${\bf E}_{L,T}(-k)=-i\omega{\bf A}_{L,T}(-k)$. (\ref{JJLT}) is just
Ohm's law where the longitudinal and transverse conductivities at finite $\omega, \,k$ are
respectively

\begin{eqnarray}
&&\sigma_L(k)=\frac{G_{L}(k)}{-i\omega}\nonumber\\
&&\sigma_T(k)=\frac{G_{T}(k)}{-i\omega}
\end{eqnarray}

For the RN-AdS black-hole, $G_L=G_{zz}$ and $G_T=G_{xx}$ in the hydrodynamic limit  $\omega/T<<1, \quad k/T<<1$ have been explicitly worked out
in~\cite{SIN2}.  We quote the results here for a physical interpretation to follow for a cold
fermionic system.  In particular

\begin{eqnarray}
&&\sigma_L(k)=i\omega\left(\frac{A_T}{\omega^2-k^2/3}+\frac{A_L}{i\omega-D_Lk^2}\right)\nonumber\\
&&\sigma_T(k)=-\left(\frac{A_T}{i\omega-D_Tk^2}-A_L\right)\nonumber\\
\label{GGLT}
\end{eqnarray}
with the real residues

\begin{eqnarray}
&&A_L=\frac{N_c^2\gamma^2}{4\pi^2}\frac 1{8b}\left(\frac{2-a}{1+a}\right)^2\nonumber\\
&&A_T=\frac{N_c^2\gamma^2}{4\pi^2}\frac{3a}{4b^2} \frac 1{1+a}\nonumber\\
\label{PARAMETERS}
\end{eqnarray}
The transverse and longitudinal diffusion constants are tied $D_L/D_T=(2+a)$ and $2D_T=b/(1+a)$.
The parameters $a$ and $b$ have been defined in section 2. They are explicitly given as

\begin{eqnarray}
&&a=\frac{8 }{3 }\frac{\gamma^2\mu^2}{\pi^2T^2}\left(1+\sqrt{1+\frac{4\gamma^2 \mu^2}{3\pi^2T^2}}\right)^{-2}\nonumber\\
&&b=\frac 1{\pi T} \left(1+\sqrt{1+\frac{4\gamma^2 \mu^2}{3\pi^2T^2}}\right)^{-1}
\end{eqnarray}

The elastic mode in(\ref{GGLT}) follows from fermion number conservation, since $\partial^\mu J_\mu=0$.
Thus $J_0=(k/\omega)J_z$ so that $G_{00}=(k^2/\omega^2)G_{zz}$. $G_{00}$ triggers scalar sound waves.
At zero density or $\mu=0$, $a=0$ and the residue $A_T=0$, hence the elastic or sound mode drops from the
longitudinal conductivity. As a result,  the DC  conductivities for $k=0, \,\omega\rightarrow 0$
are $$\sigma(0)_L=\sigma_T(0)=A_L=N_c^2\gamma^2 T/4\pi. $$  In contrast, at zero temperature $a=2$ and the residue
$A_L=0$. As a result the fermionic conductivities are

\be
\sigma_L(0)=\sigma_T(0)=i\frac{A_T}{\omega}=i\frac{N_c^2\gamma^4\mu^2}{6\pi^2\omega}
=\frac{n}{\mu}\frac1{-i\omega}
\label{DRUDE}
\ee
The fermion current is $\pi/2$ out of phase with the fermionic electric field.
(\ref{DRUDE}) is the Drude conductivity $\sigma_D=ne_F^2/m\tau$  for
{\it free fermions} with an {\it effective}
mass of order $\mu$. The collision time is $\tau\approx 1/\omega$ to this order of the approximation.
Indeed, for a harmonic fermion electric field, the free fermion velocity increases linearly with time
until the harmonically oscillating field reverses (balistic regime). This system can be characterized
by an electric-like or screening mass

\be
m_E^2=\Pi_{00}(0,\vec{0})=3\Pi_L(0,\vec{0})=\frac{3n}{\mu}=\frac{\partial n}{\partial \mu}
\ee
where the limits in the first two equalities are $\vec{k}\rightarrow \vec{0}$ and $\omega\rightarrow 0$ 
sequentially.  Here we have defined $\Pi_{\mu\nu}=G_{\mu\nu}/(-i\omega)$.

In general, near the cold fermion limit

\begin{eqnarray}
&&A_L= \frac{N_c}{12\sqrt{ 3 } \gamma}\frac{T^2}{\mu}+{\cal O}(T^3)\nonumber\\
&&A_T= \frac{N_c^2\gamma^4\mu^2}{6\pi^2}+{\cal O}(T)\nonumber\\
&&D_L={\frac1{\sqrt3 \gamma}}\frac 1{\mu} +{\cal O} (T)\nonumber\\
&&D_T=\frac14\frac1{\sqrt{3}\gamma}\frac 1{ \mu}+{\cal O}(T)
\end{eqnarray}
The longitudinal diffusive pole vanishes. For cold fermions in RN-AdS the longitudinal
channel exhibits a sound mode that propagates with the speed $1/\sqrt{3}$ which
is expected in the conformal limit.  This point is analogous to the
zero sound mode suggested recently in~\cite{SON,KIM} using D3/D7 in the probe
approximation.  The sound mode is damped at zero temperature because of the
persisting finite black hole entropy or degeneracy of the colorless fermionic clusters
as we detail below for our case.

In case there is a pairing mechanism, 
we expect the degenerate fermi liquid to be replaced by a dual repulsive bosonic liquid. The longitudinal sound wave  then should be interpreted as the superfluid wave, the zero sound.  The fermionic conductivities are replaced by the Meissner-like mass $m_M^2$, as the electric-like mass $m_E^2$ is common to both liquids. These masses follow
readily from the  constant mode effective action as well.
That is by promoting the zero temperature pressure $p$ to a Lagrange  density, 

\be
{\cal L}(A_0,\vec{A}) =\frac{\gamma^2N_c^2}{24\pi^2}
\left((\mu+A_0)^2-\vec{A}^2\right)^2
\ee 
 One can read off 

\be
m_E^2=3m_M^2=\frac{3n}{\mu}
\ee
(39) was already forseen in~\cite{KAPUSTA} by noting that $m_E^2=\partial^2p/\partial\mu^2$ even at finite
temperature. The superfluid mode velocity is just $1/\sqrt{3}$ through the longitudinal
substitution $A_\mu\rightarrow \partial_\mu\pi$
as a Goldstone mode. 
The  massless superfluid phonons in 3
dimensions are expected to contribute cubically to the specific heat, which is 
accounted for in~(\ref{SPECIFIC}) at next to next to leading order.
 
\section{Visco-Elastic Analysis}
 
In a cold fermionic system the fermionic conductivities can be analyzed using the visco-elastic equations
which are generalized elasticity equations in a medium that is characterized by bulk $K$ and shear $M$
modulii, and bulk $\xi$ and shear $\eta$ viscosities~\cite{KIM}.  Since the cold fermionic response in
holography shows no sign of elasticity in the transverse channel, ie $\sigma_T$ is purely diffusive, it
follows that in holography $M=0$. So the cold fermionic limit is liquid.  The bulk viscosity $\xi=0$
since the underlying fermionic interactions are mediated by a conformal gauge theory.  Thus, the cold
fermionic liquid in holography is characterized only by a bulk modulus $K$ and a shear viscosity $\eta$.

 In the visco-elastic framework, we expect ~\cite{KIM}

\begin{eqnarray}
&&\tilde\sigma_L(k)=i\omega\left(\frac{n/m}{\omega^2-(K/mn) k^2 +i  (4\eta/3mn)\omega k^2}\right)\nonumber\\
&&\tilde\sigma_T(k)=-\left(\frac{n/m}{i\omega-(\eta/mn) k^2}\right)
\end{eqnarray}
for massive fermions of mass $m$ and density $n$ with $M=0$ and $\xi=0$.
These constitutive equations are to be compared with (\ref{GGLT}) for $a=2$ or $A_L=0$

\begin{eqnarray}
&&\sigma_L(k)=i\omega\left(\frac{A_T}{\omega^2-k^2/3}\right)\nonumber\\
&&\sigma_T(k)=-\left(\frac{A_T}{i\omega-D_Tk^2}\right)
\label{SSLT}
\end{eqnarray}
Thus, the identification

\begin{eqnarray}
&&n/m=A_T\nonumber\\
&&K/mn=c_L^2=1/3\nonumber\\
&&\eta/mn=D_T
\label{IDENTIFY}
\end{eqnarray}
The absence of attenuation in the longitudinal conductivity in (\ref{SSLT}) is a higher order effect not retained
in the analysis in~\cite{SIN}.  The constitutive or visco-elastic equations show that this calculation is actually
not needed. The superfluid mode in (\ref{GGLT}) and (\ref{SSLT})  is attenuated through the dispersion law

\be
\omega\approx c_Lk -i\frac{2D_T}3k^2
\label{DISPERSE}
\ee
with $c_L=1/\sqrt{3}$.   This is indeed expected since $c_L^2=\partial p/\partial\epsilon=1/3$ since $p=\epsilon/3$
in black-hole holography. Also, it follows that

\be
\frac{\eta}{n}=n\frac{D_T}{A_T}=\frac 1{4\sqrt{3}\gamma}
\label{VISCOSITYRATIO}
\ee
after using $m=n/A_T=2\mu/3$ from (\ref{IDENTIFY}).  This result generalizes readily to arbitrary
temperature in holography by using the general expressions for $D_T$ and $A_T$. Specifically,

\be
\frac{\eta}n= \frac {\pi T}{8\gamma^2\mu} \left(1+\sqrt{1+\frac{4\gamma^2 \mu^2}{3\pi^2T^2}}\right)
\label{ETANGENERAL}
\ee
which is in agreement with the thermodynamical relation (\ref{ENTROPYRATIO})

\be
\frac{\eta}n=\frac{\eta}s\frac sn=\frac 1{4\pi}\frac{s}{n}
\ee
The ratio $\eta/s=1/4\pi$  holds for arbitrary $T,\mu$ for the RN AdS black-hole thanks to the finite
entropy density at $T=0$. The non-vanishing of the entropy at $T=0$ provides   
interesting example of residual entropy  as we noted earlier.  This point will be stressed further below.

The emerging physical picture suggested by thermodynamics of cold and strongly coupled fermions in holography is the following:
 At zero temperature fermions carry finite energy and internal degeneracy $f$ per particle. 
They form a  degenerate liquid (maybe a glassy liquid) with a massless sound mode  that damps because of the finite shear viscosity of the liquid caused by $f$. The latter is due to a finite 
entropy per particle carried by the clusters as encoded by the nonzero black hole entropy at zero temperature. Away from zero temperature, the clusters break into their consituent quarks  that
contribute linearly to the specific heat.  

\section{Trapped Massive Fermions as Massles Fermions}

In this paper we wish to extract
some generic ratios   that we deem general for a strongly coupled fermionic system
from RN AdS.  Such systems have been studied intensively in the laboratory using
optical and magneto-optical traps of Li$^6$ at Duke~\cite{DUKE} and K$^{40}$ at JILA~\cite{JILA}.
By sweeping magnetically through the traps, the fermionic atoms are driven through the Feshbach
resonance leading to a large S-wave scattering length and strong pair interactions (unitarity limit).

Here we would like to show that a large sample of
trapped massive  fermions in a harmonic well behave similarly to free massless and
untrapped fermions, albeit with higher degeneracy.  Indeed, consider free massive fermions in a
harmonic trap at finite temperature and density. For simplicity of the argument in this section,
we choose units with $m=T=\omega=1$. Here $\omega$ is the frequency of a symmetric  trap. For a
homogeneous trap and a large density of fermions the Thomas-Fermi approximation applies.
In particular, the number $N$ of fermions in the trap is

\be
N=\int d\vec{x}\, n(\vec{x}) =d\int d\vec{x}\,d\vec{p}
\frac 1{z^{-1}e^{\vec{p}^2/2+\vec{x}^2/2}+1}
\label{TH1}
\ee
with $z=e^{\mu}$ the fermionic fugacity.  (\ref{TH1}) follows from the Thomas-Fermi approximation
through the substitution $\mu\rightarrow \mu -V(\vec{x})$ with $V(\vec{x})=\vec{x}^2/2$ the
harmonic potential.  Straightforward algebra yields

\be
N=\frac{4dV_5}{V_2}\,\int \,d\vec{p}
\frac 1{z^{-1}e^{|\vec{p}|}+1}
\label{TH2}
\ee
While (\ref{TH1}) describes massive fermions in a harmonic trap of degeneracy $d$, its analogue
(\ref{TH2})  describes massless and untrapped or free fermions of degeneracy $4dV_5/V_2>d$. Here
$V_N$ is the volume of the $S^N$  unit sphere.  This relation holds for arbitrary space dimensions.
We note that the degeneracy is $T$ and $m$ dependent when we restore the scales, but that does not
affect the Gibbs relations since in the Thomas-Fermi approximation all bulk thermodynamics follows
from relations of the type (\ref{TH1}).

It follows immediately that the energy per particle and entropy per particle of trapped but free massive
fermions and untrapped massless fermions are identical in 3 space dimension for a sufficiently large
and homogeneous number of particles to allow for the Thomas-Fermi approximation. In particular,

\begin{eqnarray}
&&\frac{E}{NT}=3\frac{f_4(z)}{f_3(z)}\nonumber\\
&&\frac{S}{N}=4\left(\frac{f_4(z)}{f_3(z)}-\frac 14 {\rm ln} z\right)
\label{TH3}
\end{eqnarray}
after we have restored $T$ for dimensions with

\be
f_n(z)=\frac 1{\Gamma(n)}\int_0^\infty\, dx\,\frac {x^{n-1}}{z^{-1}e^x+1}
\label{TH4}
\ee
In the cold limit

\begin{eqnarray}
&&\frac {E}{N\mu}=\frac 34 +\frac{\pi^2}{2}\left(\frac{T}{\mu}\right)^2+{\cal O}(T^3)\nonumber\\
&&\frac{S}{N}=\pi^2\left(\frac{T}{\mu}\right)+{\cal O}(T^2)
\label{TH5}
\end{eqnarray}
for {\it both} untrapped  massless and harmonically trapped massive fermions.  The occurence of
$3/4$ in the massive case for the ratio $E/N\mu$ already signals the conformal transmutation
due to the harmonic well in the Thomas-Fermi approximation. This point is made clear in (\ref{3over4}).
Note that by considering ratios in (\ref{TH3}) all the degeneracy mismatch between massless
and massive got absorbed.

This remarkable thermodynamical transmutation between free massless and massive but
harmonically trapped fermions, allow us to suggest that the present holographic approach
with untrapped and massless fermions at strong coupling, is actually well suited for describing
trapped fermionic atoms in the unitarity limit. In particular ratios of the type (\ref{TH3}) and their
expanded version (\ref{TH5}) may capture the essential model independent physics. We now
proceed to compare our holographic results with current atomic experiments.

\section{Atomic data from Holography}

Since we found that {\it free} cold trapped atoms behave like massless fermions, we are tempted to map
the thermodynamics of the {\it interacting} cold trapped atoms to that of the RN AdS black hole which is
relevant for massless and strongly interacting massless fermions.
To compare the black hole results with those from atomic experiments, we introduce an
interaction parameter $\xi$
\be
\sqrt{\xi}=\mu/k_F=\mu/T_F 
\ee
to characterize  the degree of interaction in the  ground state.
Here,  $T_F=E_F=k_F$ is the Fermi temperature for non-interacting 
massless fermions. The parameter $\xi$ is 
used for cold fermionic systems but we want to define 
it in general for strongly interacting fermions like our
dual RN AdS black hole with bulk filling branes. 
For that we recall that for strongly interacting fermions, the energy per 
particle normalized to the Fermi momentum is defined as 

\be
\frac{E}{NE_F}=\frac{\epsilon}{n\mu}\frac{\mu}{k_F}=\frac 34 \sqrt{\xi} 
\ee 
To determine $\xi$ in terms of the gravity and brane  parameters, we note  that the 
fermion number is unaffected by the interaction. 
Consider first the noninteracting fermion system, where the density is

\be n_{free}=d k_F^3/6\pi^2  \;\;\; with \;\;\; d=N_fN_c 
\ee 
for spinless  fermions with $N_f$ flavor (in the fundamental representation of color and flavor).  
As we turn on the strong (attractive) interaction, we expect the fermions to bind into composites and the Fermi energy $\mu$ 
to depart  from its free value $k_F$.  The number density of the interacting system is given in our holographic model by 
\be
n_{int}=\frac{N_c^2\gamma^4}{6\pi^2}\mu^3.
\ee
We take the parameter determined from the $D3/D7$ system 
to set   $\gamma=(N_f/N_c)^{1/2}$.   Since fermion number is unchanged, we find that the 
fundamental parameter $\gamma$ in holography ties with the fundamental parameter $\xi$
in cold fermion systems through~\footnote{We note that this is the only time where we make explicit use of the 
parameters set of the D3/D7 embedding in~\cite{SIN}. Without it, $\xi=\left({d}/{N_c^2\gamma^4}\right)^{2/3}.$ }

\be
\xi=  \gamma^{-4/3}
\ee 
Now we want to map the thermodynamics of a
cold fermion system to that of the charges in bulk filling branes discussed in section 3 and 4. 
 Atomic experiments in the unitarity limit  give $\xi=0.41$~\cite{LUO,KINAST}
 to fit the ground state energy.
We will use this empirical value in our numerical estimates below. 

The
entropy per particle in holography is nonzero at the coldest point
\be
\frac{S}{N}=\frac{\pi}{\sqrt{3}}\,\xi^{3/4}\approx  0.93 .
\label{COLDENTROPY1}
\ee
   Recent atomic physics experiments~\cite{LUO} have measured
the entropy of trapped cold atoms by using magnetic field sweeps across the trap. The reported
measurements reveal a coldest entropy per particle of about 1 that is consistent with our estimate.
The entropy (\ref{COLDENTROPY1}) is at the origin of a finite cold viscosity per particle

\be
\frac{\eta}{n}=\frac{\xi^{3/4}}{4\sqrt{3}}\approx 0.07
\ee
which we note to be larger than $1/6\pi$ as conjectured in~\cite{GELMAN}.  At this point, it
is worth noting that the atomic measurements of $\eta/n=<\alpha>$ in~\cite{LUO}  at their
coldest point are also consistent with our holographic estimate.   If one  extrapolates the   entropy to be zero at zero temperature, it suggests  $\eta \sim s /4\pi  \sim 0$ 
in the ground state in the unitarity limit, which is not consistent with $\eta/s=1/4\pi$.  Current atomic measurements of the entropy and/or
viscosity at the coldest point in the unitarity limit offers a direct measurement of the cold
black hole entropy through holography.   
At finite temperature, the entropy per particle is

\be
\frac{S}{N}=\frac{\pi^2}{2}\xi  \,x\left(1+\sqrt{1+\frac{4 \xi^{-1/2}}{3\pi^2 x^2}}\right)
\ee
with $x=T/T_F$.  At high temperature the entropy per particle is $S/N\approx \pi^2 x$. Similarly, the
energy per particle is  

\be
\frac{E}{NT_F}=\frac{\pi^2}{16}\,\xi \,x^2
\left(\left(1+\sqrt{1+\frac{4\xi^{-1/2}}{3\pi^2 x^2}}\right)^2+\frac{8\xi^{-1/2}}{3\pi^2 x^2}\right)
\ee
To make the comparison with atomic experiments manifest, we set $E_0$ to be the value
of $E$ at zero temperature or $x=0$, and defined the shifted energy $\Delta=(E-E_0)/NT_F$
which is zero for $x=0$. Thus

\be
\Delta=\frac{E-E_0}{NT_F}=\frac{\pi^2}{16}\,\xi \,x^2
(\left(1+\sqrt{1+\frac{4\xi^{-1/2}}{3\pi^2 x^2}}\right)^2-\frac{\xi^{1/2}}{12}
\label{DELTA}
\ee
$\Delta$ is the analogue of $(E_{840}-E_0)/{E_F}$ used in the empirical analysis in~\cite{LUO}.\footnote{840 here is the magnetic field in Gauss unit to confine the fermions.}
This translates to

\be
\Delta\approx 0.51x\left(x+\left(x^2+0.21\right)^{0.5}\right)
\ee
which is to be compared with $\Delta\approx 2.64\,x^{1.43}$ empirically in the
nonsuperfluid phase with $x>T/T_F\approx 0.34$~\cite{LUO}.

%
%
\begin{figure}[]
  \begin{center}
    \includegraphics[width=10cm]{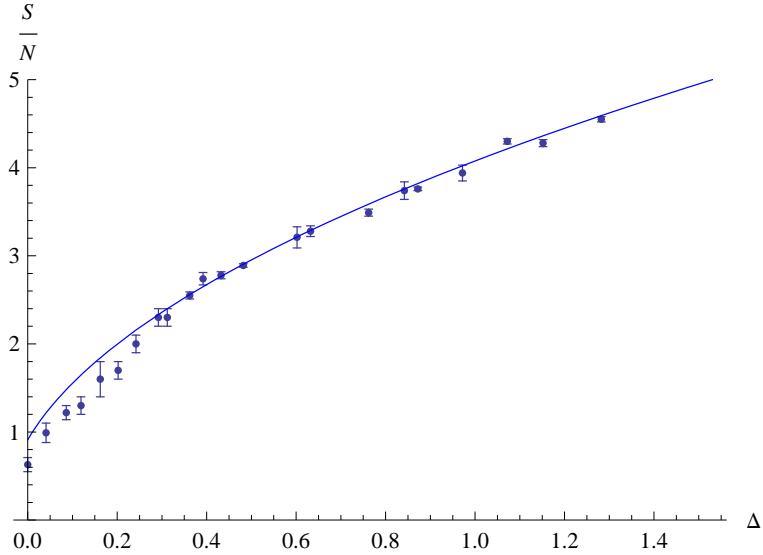}
  \caption{$S/N$ versus energy $\Delta$: Holography (solid line: $\xi=0.4$) and
                  atomic data~\cite{LUO}.}
  \label{Fig:F2}
  \end{center}
\end{figure}

In terms of the shifted energy (\ref{DELTA}), the entropy per particle simplifies

\begin{eqnarray}
\frac{S}{N}=(2\pi {\xi^{1/2}})\left({\Delta +\frac{ {\xi^{1/2}}}{12}}\right)^{1/2}
\approx 4.02\,({\Delta +0.05})^{0.5}
\end{eqnarray}
which is to be compared with

\be
\frac{S}{N}=(4.0\pm 0.2)\,\Delta^{0.45\pm 0.01}
\ee
obtained empirically in~\cite{LUO,KINAST} in the nonsuperfluid regime or $E/T_F>0.94\pm 0.05$.
The holographic energy dependence of the entropy compares remarkably with the atomic
experiments as we show in Fig.~\ref{Fig:F2}.   We note that the small temperature corrections in (\ref{TH5})
for free massless
fermions give at low temperature $S/N\approx \pi\sqrt{2\Delta}\approx 4.44\,\Delta^{0.5}$ which
overshoots the data.  The corresponding viscosity per particle reads

\be
\frac{\eta}{n}
\approx 0.32 \,({\Delta +0.05})^{0.5}
\ee

\section{Conclusions}

We have used an holographic construction to analyze cold and hot massless fermions including
fermionic back-reaction as discussed recently by one of us~\cite{SIN}.  The ensuing gravitational metric
is RN AdS black-hole as opposed to  AdS black hole for the probe approximation. Holography allows
a number of predictions for the bulk thermodynamics and transport near the cold point, that are readily
amenable to testing against current atomic experiments involving ${\rm Li}^6$ or ${\rm K}^{40}$ atoms near the
so called Feshbach resonance.

We have shown that holography supports a finite entropy and thus shear viscosity at the coldest point.
While the D3/D7 embedding allows for a derivation of the fermionic ground state energy per particle, the
result depends only on $\sqrt{\xi}=(N_c/N_f)^{1/3}$, a fundamental parameter in holography.  We recall
that trapped atomic experiments put $\xi=0.41$. We have found that the effects of temperature as suggested
by the RN AdS black hole construction compare favorably with current atomic measurements. In particular
the entropy versus energy dependence in cold atoms is derived acuratly. Given the simplicity of the theoretical
construction this is certainly very pleasing.

The idea of a minimal cold viscosity per particle in trapped atoms was initially suggested in~\cite{GELMAN}
who argued that $\eta/n\geq 1/6\pi\approx 0.05$.  It also follows from a quasi-normal mode analysis
using D3/D7 in the probe approximation~\cite{SON,KIM}.  In particular $\eta/n=1/4$~\cite{KIM}.   Our analysis
differs from~\cite{SON,KIM} in that the full gravity back-reaction due to the presence of the fermions is considered. Our
construction also shows explicitly how the shear viscosity ties with the entropy (degeneracy) of the cold black
hole at zero temperature.

The specific heat and equation of state $p=\epsilon/3$  we discussed in this paper strongly suggest that the relevant degree of freedom at low temperature is 
massless and fermionic.
Very recently another bosonic system is reported to show  the thermodynamic  
and transport behavior of fermi liquid \cite{Gubser:2009qt}.  
However, the conductivity of charged ads BH   at a cold point is $\sim T^2$ 
while we expect   $1/T^2$ for fermi liquid. Therefore   the issue of
gravity dual of fermionic system is not totally clear and it is still waiting further investigation.  

  The existence of a finite
entropy at zero temperature for charged black holes is very intriguing as no Bekenstein-Hawking radiation
can be assigned to it in standard gravity.  Precise measurements of the entropy of trapped cold atoms may
shed light on this fundamental concept at the coldest point. The latter is currently probed only by theoretical
extrapolations to have zero entropy at $T=0$. We have shown that these extrapolations conflict with
the finite shear viscosity quoted in~\cite{LUO} at the coldest point.

\vskip 1cm

{\bf Note added:} 
After submitting the first version of this paper to the archive,  Ref. ~\cite{rey} was brought to our attention
regarding some overlap with sections 3,4.  However, our results are different.

\section{Acknowledgments}
IZ thanks Hanyang university for hospitality.
This work was supported in part by the WCU project of Korean Ministry of Education,
Science and Technology (R33-2008-000-10087-0).
The work of IZ  was supported in part by US-DOE grants
DE-FG02-88ER40388 and DE-FG03-97ER4014.
The work of SJS was supported in part by KOSEF Grant R01-2007-000-10214-0   and
 by the National Research Foundation of Korea(NRF) grant funded by the Korea government(MEST) (No.20090063068).

\end{document}